\documentclass[aps,preprint,superscriptaddress,footinbib,floatfix,showpacs]{revtex4} 
\usepackage{graphicx}

\begin{document} 
  
\title{The Semiclassical Regime of the Chaotic Quantum-Classical Transition} 
      \vbox to 0pt{\vss 
                    \hbox to 0pt{\hskip-40pt\rm LA-UR-03-9308\hss} 
                   \vskip 25pt} 
\preprint{LA-UR-03-9308} 
\author{Benjamin D. Greenbaum} 
\affiliation{Department of Physics, Columbia University, New York,
New York 10027} 
\author{Salman Habib} 
\affiliation{T-8, Theoretical Division, The University of
California, Los Alamos National Laboratory, Los Alamos, NM 87545} 
\author{Kosuke Shizume} 
\affiliation{Institute of Library and Information Science, University
of Tsukuba, 1-2 Kasuga, Tsukuba, Ibaraki 305-8550, Japan} 
\author{Bala Sundaram} 
\affiliation{Graduate Faculty in Physics \& Department of Mathematics,
City University of New York - CSI, Staten Island, New York 10314} 

\begin{abstract}
  
{An analysis of the semiclassical regime of the quantum-classical
transition is given for open, bounded, one dimensional chaotic
dynamical systems. Environmental fluctuations -- characteristic of all
realistic dynamical systems -- suppress the development of fine
structure in classical phase space and damp nonlocal contributions to
the semiclassical Wigner function which would otherwise invalidate the
approximation. This dual regularization of the singular nature of the
semiclassical limit is demonstrated by a numerical investigation of
the chaotic Duffing oscillator.}

\end{abstract}
\maketitle

{\bf The process by which a classical dynamical system emerges as a
sufficient approximation to a quantum dynamical system has been a
major topic of discussion since the inception of quantum
mechanics. The singular nature of the semiclassical limit
($\hbar\rightarrow 0$)~\cite{berry1} lies at the center of this
debate. Classical behavior cannot emerge as the smooth limit of a
closed quantum system with a nonlinear Hamiltonian as these classical
evolutions violate the unitary symmetry of quantum
mechanics~\cite{1}. Moreover, the symplectic geometry of a chaotic
classical phase space generates infinitely fine structures which the
uncertainty principle prevents a quantum dynamical system from
tracking as $t\rightarrow\infty $ ~\cite{2}. Thus the pathologies
associated with the semiclassical limit are most dramatic in
classically chaotic dynamical systems. In these systems, small $\hbar$
expansions of physical averages have been shown to fail at a finite
time \cite{Zas}. This work addresses how the incompatibility of
quantum dynamics and classical chaos can be resolved in the framework
of open quantum and classical systems.  In particular, the stability
of the semiclassical limit is recovered by an environment-induced
coarse graining of both classical and quantum dynamics, from which a
threshold condition is derived.}

As all realistic systems interact with their environment, the modern
approach to understanding the quantum-classical transition (QCT)
relies on the open system paradigm. In this paradigm, the dynamical
system is not considered in isolation but analyzed taking its external
interactions into account. These {\em environmental} interactions are
of two types depending on whether it is possible to make measurements
on the environment or not. If the environment is in principle
unobservable, then the system is described by the reduced density
matrix obtained from the full system-environment density matrix and
tracing over the environment. If, on the other hand, certain
measurements are possible on the environment, then the resulting
reduced density matrix for the system depends on the results of these
measurements. System evolution in this second case is therefore said
to be {\em conditioned} on the observation results~\cite{cm,car}. The
conditioned system state, as it evolves, is said to define a {\em
quantum trajectory} and inequalities governing the existence of a
classical trajectory limit of quantum trajectories in continuously
measured quantum systems have now been obtained
~\cite{SalKurtTan}. Since it yields effectively classical trajectories
-- from a single realization of a measurement record -- we will call
this pathway the {\em strong} form of the quantum-classical
transition.

If the environment is not amenable to observation, or if one decides
to throw away the results of measurements on the environment -- which
amounts to the same thing -- then the evolution of the reduced density
matrix of the system is given by an unconditioned master equation, as
one must average over all possible measurement outcomes. It is now no
longer possible to obtain the classical trajectory limit as discussed
above.  One must now compare quantum and classical distributions (or,
equivalently, the underlying moment hierarchy) against each other:
This constitutes the {\em weak} form of the quantum-classical
transition. For any given situation, if the inequalities defining the
strong form of the QCT are satisfied, then the weak form of the QCT
immediately follows. The reverse is not true, however.  Finally, it
should be noted that there is no direct connection between the
inequalities defining the strong form of the QCT and the ones defining
the weak form derived here.

The weak version is just another way to state the conventional
decoherence idea~\cite{deco}; however, as discussed in Ref.~\cite{1},
mere suppression of quantum interference does not guarantee the QCT
even in the weak form. To address this problem, our purpose here is to
present a semiclassical analysis of the QCT for bounded, classically
chaotic open systems focusing on the {\em regularization} of the
singular $\hbar\rightarrow 0$ limit via the weak form of the
environmental interaction. This is distinct from the state {\em
localization} characteristic of the strong form of the QCT. Given a
small, but finite, value of $\hbar$, we aim to establish the existence
of a timescale beyond which the dynamics of open quantum and classical
systems becomes statistically equivalent if the environmental
interaction is sufficiently strong.

It is important to keep in mind that the results obtained in this work
are intrinsically different from the inequalities describing the
strong form of the QCT given in Ref.~\cite{SalKurtTan}. The analysis
here refers to the {\em coarse-grained} distribution function
(averaging over all measurements), whereas the analysis in
Ref.~\cite{SalKurtTan} refers to the {\em fine-grained} distribution
for a single measurement realization. The strong form of the QCT
requires that the quantum distribution be localized, and this is
enforced by the conditions, for weak and strong nonlinearity,
respectively \cite{SalKurtTan}:
\begin{eqnarray} && 8k\gg\sqrt{\frac{(\partial_x^2 F)^2|\partial_x
F|}{2mF^2}} \nonumber\\ 
&& 8k\gg\frac{(\partial_x^2 F)^2\hbar}{4mF^2}.
\end{eqnarray}
In addition, the key low-noise condition on the trajectories is
enforced by a double-sided inequality \cite{SalKurtTan}:  
\begin{equation}
\frac{2|\partial_x F|}{s}\ll\hbar k\ll\frac{|\partial_x F|s}{4},
\label{newtonq}
\end{equation}

In the equations above, where $k$ is the measurement strength, $\eta$
is the measurement efficiency, $m$ is the particle mass, $F$ is the
force on the system measured at the expectation value
$\langle\hat{X}\rangle\equiv x$ and $s$ is the typical action of the
system in dimensionless units. This second inequality first states
that if the measurement is too weak, the measurement signal is noisy
as it results from sampling a wide distribution. As the measurement
strength is increased, the distribution becomes sharply peaked and
there is very little noise in the measurement result. However, if the
measurement is too strong, the resulting backaction drives the
dynamics too hard and the trajectory becomes noisy. Either of these
limits need not prevent a weak QCT from existing as the weak version
of the QCT is not a direct consequence of state localization and
therefore can exist even when these inequalities are not satisfied. It
does not matter if the coarse-grained distribution is too wide, as
long as the classical and quantum distributions agree, and, even if
the backaction noise is large, the coarse-grained distribution remains
smooth and the weak quantum-classical correspondence can still exist.

We demonstrate that, for a bounded open system with a classically
chaotic Hamiltonian, the weak form of the QCT is achieved by two
parallel processes. First, the semiclassical approximation for quantum
dynamics, which breaks down for classically chaotic systems due to
overwhelming nonlocal interference, is recovered as the environmental
interaction filters these effects. Second, the environmental noise
restricts the foliation of the unstable manifold, the set of points
which approach a hyperbolic point in reverse time, allowing the
semiclassical wavefunction to track this modified classical
geometry. Our approach explicitly incorporates both the stretching and
folding typical of hyperbolic regions and the role of the environment
as a filter on a phase-space quantum distribution. Thus our results
are different from the purely local analysis of Ref.~\cite{zp} and the
study of decay of off-diagonal coherences in Ref.~\cite{Kos} which
focuses on the filtering aspect of the QCT.

We examine a simple model of a quantum system weakly coupled to the
environment so as to maintain complete positivity for the subsystem
density matrix, $\hat{\rho}(t)$, while subjecting it to a unitarity
breaking interaction.  These conditions mathematically constrain the
master equation to be of the so-called Lindblad form ~\cite{Lind}. If
this environmental interaction couples to the position, as is often
the case, the master equation will take the form:
\begin{equation}
\frac{d\hat{\rho}}{dt}=-\frac{i}{\hbar}[\hat{H},\hat{\rho}]-
k_{env}[\hat{X},[\hat{X},\hat{\rho}]],  
\label{me1}
\end{equation}
for environmental coupling strength $k_{env}$, subsystem Hamiltonian
$\hat{H}$ and position operator $\hat{X}$
~\cite{cm,Caldeira,djj}. We have neglected the dissipative
environmental channel and kept the diffusive channel for two reasons:
(i) the coupling to the environment is always assumed to be weak and
the dissipative timescales are hence very long, longer than the
dynamical timescales of interest, (ii) the weak form of the QCT arises
only from the diffusive channel, hence dissipative effects are not of
interest here.

The master equation is often examined by taking the Wigner transform
of the density matrix in its position representation.  This yields the
Wigner function, which is a quasiprobability distribution over phase
space whose evolution can then be compared to the corresponding
classical phase space distribution function~\cite{Wig}. The Wigner
representation of the density matrix is given by
\begin{equation}
f_{w}(q,p,t)=\frac{1}{2\pi\hbar}\int_{-\infty}^{\infty}dX
e^{-ipX/\hbar}\rho(q+\frac{X}{2},q-\frac{X}{2},t).  
\label{fwdef}
\end{equation}
In this representation, the master equation for the open Wigner
evolution becomes 
\begin{equation}
\frac{\partial f_{w}}{\partial t} =
L_{cl}f_{w}+L_{q}f_{w}+D\frac{\partial^{2}f_{w}}{\partial p^{2}},
\label{wme} 
\end{equation}

where the classical Liouville operator $L_{cl}\equiv
-(p/m)\partial_{x}+ (\partial_{x}V)\partial_{p}$ and the quantum
Liouville operator $L_{q} \equiv\sum_{n\geq
1}(\hbar^{2n}(-1)^{n}/(2^{2n}(2n+1)!))~\partial_{x}^{2n+1}V
\partial_{p}^{2n+1}$ and $D=\hbar^{2}k_{env}$.  When $L_{q}=0$, this
equation reverts to the classical Fokker-Planck equation. It is
important to keep in mind that the specific form of the diffusion
coefficient (alternatively, $k_{env}$) depends strongly on the
physical situation envisaged. Thus, if the master equation describes a
weakly coupled, high temperature environment, $D=2m\gamma k_BT$
($\gamma$ is the damping coefficient)~\cite{Caldeira}, whereas for a
weak, continuous measurement of position, the diffusion due to quantum
backaction is $D=\hbar^2 k$~\cite{cm}. The results in this paper hold
for all of these cases.

The closed-system evolution ($k_{env}=0$) of the quantum and classical
Liouville ($L_q=0$) equations for nonlinear systems yield very
different results. The Wigner evolution is quickly dominated by
nonlocal interference effects and has rapidly oscillating positive and
negative components, whereas the classical evolution is always
positive but develops very fine-scale structure. Numerical studies for
a class of classically chaotic Hamiltonians carried out for small, but
finite $k_{env}$, show that diffusion rapidly filters fast, nonlocal
interference terms, but it also filters the fine-scale classical
structure~\cite{HabibDuff,carvalho}. Thus, the QCT connects the
quantum master equation to the classical Fokker-Planck equation and
not to the classical Liouville equation. Establishing the general
conditions under which this occurs is the main aim of this paper.
 
Once the QCT occurs, the effects of $L_q$ in the evolution specified
by Eqn.~(\ref{wme}) are subdominant. Therefore, to understand how
environmental noise acts in this limit, it suffices to consider the
behavior of the corresponding classical Fokker-Planck equation. To do
this, it is convenient to examine the underlying Langevin equations
for noisy trajectories that unravel the evolution of the classical
distribution function when $L_{q}=0$.  These are given by:
\begin{equation}
dq=\frac{p}{m} dt
\end{equation}
and
\begin{equation}
dp=f(q)dt+\sqrt{2D}dW
\end{equation}

where $f(q)=-\partial V(q)/\partial q$ and $dW$ is the Wiener measure
[$(dW)^{2}=dt$]. For constant $D$, one can just as well write
$\sqrt{2D}dW=\xi(t) dt$, where $\xi(t)$ is a rapidly fluctuating
Langevin force satisfying $\langle\xi(t)\rangle=0$ and
$\langle\xi(t)\xi(t')\rangle=2D\delta(t-t')$ over noise averages.  To
examine the effects of this force on the formation of the stable and
unstable manifolds associated with a chaotic system we perform a
perturbative expansion about a hyperbolic fixed point $(q_{eq},0)$,
where $f(q_{eq})=0$, using small-noise perturbation theory (i.e.,
assuming the effects of noise forces are small relative to the
systematic forces on dynamically relevant timescales)~\cite{Gard}. In
the leading order approximation, we can separate the dominant
systematic components from the noisy components via $q(t)\approx
q_{C}(t)+ q_{N}(t)$ and $p(t)\approx p_{C}(t)+ p_{N}(t)$, leading to
the usual Hamilton's equations for $q_{C}$ and $p_{C}$, and to the
coupled equations $dq_{N}=p_{N} dt/m$ and
$dp_{N}=m\lambda^{2}q_{N}dt+\xi dt$, where $m\lambda^2=\partial
f(q_{eq})/\partial q$ defines the local Lyapunov exponent, $\lambda$.
These have the solution~\cite{Kamp},
\begin{eqnarray}    
q(t)& = & q_{eq}+C_{+}e^{\lambda t}+C_{-}e^{-\lambda t} \nonumber\\ 
&& +\frac{1}{2m\lambda}\int_{0}^{t}du \xi(u)
\left(e^{\lambda(t-u)}-e^{-\lambda(t-u)}\right), 
\label{lansol}
\end{eqnarray}
with an analogous expression for $p(t)$. Position and momentum are then 
rescaled according to $q'=\sqrt{\lambda m}q$ and $p'=p/\sqrt{\lambda
m}$. This gives both dynamical variables the identical units of the
square root of phase space area and renders the stable and unstable
directions orthogonal. Projecting the solutions for $q'$ and $p'$
along the stable~(-) and unstable~(+) directions, we find,     
\begin{eqnarray}      
&&u_{\pm}(t)={\sqrt{2}}(q' \pm p')\nonumber\\
&&=\sqrt{2\lambda m} C_{\pm}e^{\pm\lambda t}\pm\frac{1}{\sqrt{2 \lambda m}}
\int_{0}^{t}du\xi(u)e^{\pm\lambda(t-u)}.
\label{proj}
\end{eqnarray}
The effects of noise can now be analyzed on the distribution functions
generated by the noisy trajectories.  The average over all noisy
realizations of the displacement in each direction is given by
$\langle u_{\pm}\rangle=\sqrt{2\lambda m}C_{\pm}e^{\pm\lambda t}$, as
expected from a perturbation in the neighborhood of a hyperbolic fixed
point.  When examining the second order cumulants ones sees that,
while the stable and unstable directions have variances of $\pm
\frac{D}{2m{\lambda}^{2}} (e^{\pm 2\lambda t}-1)$, the off-diagonal
cumulant is $\langle u_{+}u_{-}\rangle-\langle u_{+}\rangle\langle
u_{-}\rangle=-Dt/(m\lambda)$.  In forward time, where the evolution of
a trajectory is determined by the unfolding of the unstable manifold,
this indicates that, as the trajectory evolves, it will simultaneously
smooth over a transverse width in phase space of size
$\sqrt{{Dt}/(m{\lambda})}$.  

The above analysis implies a termination in the development of new
phase space structures at some finite time $t^{*}$, whose scaling
behavior can be determined, which need not be true in a non-compact
phase space. The average motion of a trajectory is identical to its
deterministic motion, so at time $t$, if the initial length in phase
space is $u_{0}$, its current length will be approximately
$u_{0}e^{\bar{\lambda} t}$ as its forward time evolution will be
dominated by its component in the unstable direction.  Here
$\bar{\lambda}$ is the time-averaged positive Lyapunov exponent.  If
the region is bounded within a phase space area $A$, the typical
distance between neighboring folds of the trajectory is given by
\begin{equation}
l(t)\approx {A\over u_{0}}e^{-\bar{\lambda}t},
\label{dist}
\end{equation}
where one should recall that $l(t)$ still carries the units of the
square root of phase space area.  However, since phase structures can
only be known to within the width specified above, the time at which
any new structure will be smoothed over is defined by 
\begin{equation} 
l(t^{*})\approx \sqrt{\frac{Dt^{*}}{m \bar{\lambda}}}.  
\label{nscale}
\end{equation}
The above two equations can be used to determine $t^*$, which is only
weakly dependent on $D$ and the prefactor in Eqn.~(\ref{dist}). Due to
the smoothing, one does not see an ergodic phase space region, but one
in which the large, short-time features which develop prior to $t^{*}$
are pronounced and the small, long-time features which develop later
are smoothed over by the averaging process.  Therefore, to approximate
noisy classical dynamics, a quantum system need not track all of the
fine scale structures, but only the larger features which develop
before the production of small scale structures terminates. (This
classical suppression of structure parallels recent studies of chaotic
advection-diffusion, where the efficiency of mixing is
suppressed~\cite{Voth}.  There, as here, the microscopic dynamics
generates structure on increasingly smaller scales until diffusion
terminates further development.)

The conditions under which quantum dynamics can track this modified
phase space geometry can be examined by looking at the semiclassical
Wigner function in the presence of an environmental interaction.  As
has been appreciated for a long time, the Wigner representation
provides a useful tool for semiclassical analysis~\cite{BH}.  By
examining the semiclassical Wigner function, rather than the
semiclassical wavefunction, the breakdown of the semiclassical
approximation for chaotic systems can be clearly associated with a
geometric picture. A general mixed state is an incoherent
superposition of pure state Wigner functions, where an individual
semiclassical pure state Wigner function can be formed by substituting
the usual Van-Vleck semiclassical wavefunction in Eqn.~(\ref{fwdef}).
If we allow $q$ to be perturbed by noise we can rewrite the classical
action $S(q,t)\approx S(q_{C},t)-\sqrt{2D}\int_{0}^{t} dt
\xi(t)q_{C}(t)$, as in Ref.~\cite{Kos}.  Following Berry~\cite{BH}, we
rewrite the action for the $i$-th solution to the Hamilton-Jacobi
equation as $S_{i}(q_{C},t)=\int_{q_{C}(0)}^{q_{C}(t)}dq'p_{i}(q',t)-
\int_{0}^{t}dt'H(q_{C}(0),p_{i}(q_{C}(0),t')\left(\equiv
\int_{0}^{t}dt' {\cal H}_i(t')\right)$, where $p_{i}(q,t)$ is the
$i$-th branch of the momentum curve for a given $q$. If we average
over all noisy realizations, after separating the contributions from
identical branches, the following suggestive expression for the noise
averaged semiclassical Wigner function obtains:
\begin{eqnarray} &&\frac{1}{2\pi\hbar}\int_{-\infty}^{\infty}dX
\exp\left(-{DtX^{2}\over 2\hbar^2}\right)
\bigg(\sum_{i}{\cal J}_{ii}\times \nonumber\\
&&\exp\bigg[\frac{i}{\hbar} 
\bigg\{\int_{\bar{q}_{-}}^{\bar{q}_{+}}dq'p_{i}(q',t)-pX\bigg\}\bigg]+
\nonumber\\
&&{2i}\sum_{i<j}{\cal J}_{ij}\sin\bigg[\frac{1}{\hbar}
\bigg\{\int_{q_{C}(0)}^{\bar{q}_{+}}dq'p_{i}(q',t)-
\int_{q_{C}(0)}^{\bar{q}_{-}}dq'p_{j}(q',t)\nonumber\\
&&-\int_{0}^{t}dt'\left({\cal H}_i-{\cal H}_j\right) +
\phi_{i}-\phi_{j}\bigg\}\bigg]\bigg);\\ 
\label{semiwig}
&&{\cal J}_{ij}\equiv
{C_{i}(\bar{q}_{+},t)C_{j}(\bar{q}_{-},t)\over 
\sqrt{|J_{i}(\bar{q}_{+},t)||J_{j}(\bar{q}_{-},t)|}}
\end{eqnarray} 
for Jacobian determinant $J_{i}(q,t)$ and transport coefficient
$C_{i}(q,t)$; $\bar{q}_{\pm}\equiv q\pm \frac{X}{2}$ and
$\phi_{i}=\pi\nu_{i}$, where $\nu_{i}$ is the $i$-th Maslov
index~\cite{Maslov}. 

One can analyze the dominant contributions to the above integrals
using the stationary phase approximation~\cite{2}.  If $D=0$, these
would contribute phase coherences at values of $X$ which satisfy
$p_{i}(q+X/2,t)+p_{i}(q-X/2,t)-2pX=0$ for the first term in the
summation and $p_{i}(q+X/2,t)+p_{j}(q-X/2,t)-2pX=0$ for the second
term, the former being the famous Berry midpoint rule.  For a chaotic
system, Berry argued that, due to the proliferation of momentum
branches, $p_{i}(q,t)$, arising from the infinite number of foldings
of a bounded chaotic curve as $t\rightarrow\infty$, a semiclassical
approximation would eventually fail since the interference fringes
stemming from a given $p_{i}$ could not be distinguished after a
certain time from those emanating from the many neighboring
branches~\cite{2}.  While the precise value of this time has since
been challenged numerically, the essential nature of this physical
argument has remained valid~\cite{HellTom}.  However, the presence of
noise acts as a dynamical Gaussian filter, damping contributions for
any solutions to the above equation which are greater than
$X\approx\hbar/\sqrt{Dt}$.  In other words, noise dynamically filters
the long ``De Broglie'' wavelength contributions to the semiclassical
integral, the very sort of contributions which generally invalidate
such an approximation.  If we rescale the above result and combine it
with our understanding of how noise effects classical phase space
structures, we can qualitatively estimate whether or not a
semiclassical picture is a valid approximation to the dynamics.  As
discussed earlier, $t^*$ is the time when the formation of new
classical structures ceases and $l(t^*)$ is the associated scale over
which classical structures are averaged. The key requirement is then
that the semiclassical phase filters contributions of size
\begin{equation}
{\sqrt{\bar{\lambda} m}\hbar\over\sqrt{Dt^{*}}}\lesssim l(t^{*}).
\label{result}
\end{equation}
In other words, for a given branch, the phases with associated
wavelengths long enough to interfere with contributions from
neighboring branches will be strongly damped, and the intuitive
semiclassical picture of classical phase distributions decorated by
local interference fringes will be recovered. 

The weak form of the QCT is completed when the inequality
(\ref{result}) is satisfied. Substituting the scale of
classical smoothing (\ref{nscale}) in this inequality, we find
\begin{equation}
Dt^{*}\gtrsim {\bar{\lambda} m \hbar}.
\label{tstar}
\end{equation}
[Note that the purely classical quantity $t^*$ is first independently
determined by solving Eqn.~(\ref{nscale}) and then compared to the RHS
of the above equation.] While, the lefthand side of the inequality
contains the mutually dependent $t^*$ and $D$, the right hand side
depends only on fixed properties of the system and $\hbar$.  This
condition therefore defines a threshold at which the semiclassical
approximation becomes stable and which may be set in terms of either
$D$ or $t^*$.  Once the threshold is met, $t^*$ becomes the time beyond
which the semiclassical description is valid.  The semiclassical nature
of this condition becomes more evident on defining $S=l(t^{*})^2$
which, given that $l^2$ is an areal scale in phase space for the
diffusion averaged dynamics, has dimensions of action.  A physical
interpretation is more apparent on rewriting (\ref{tstar}) as
$S=l(t^*)^2 \gtrsim \hbar$, which is readily identified as
the usual condition for the validity of a semiclassical analysis. 
\begin{figure}[htbp]
\includegraphics[width=3.7in]{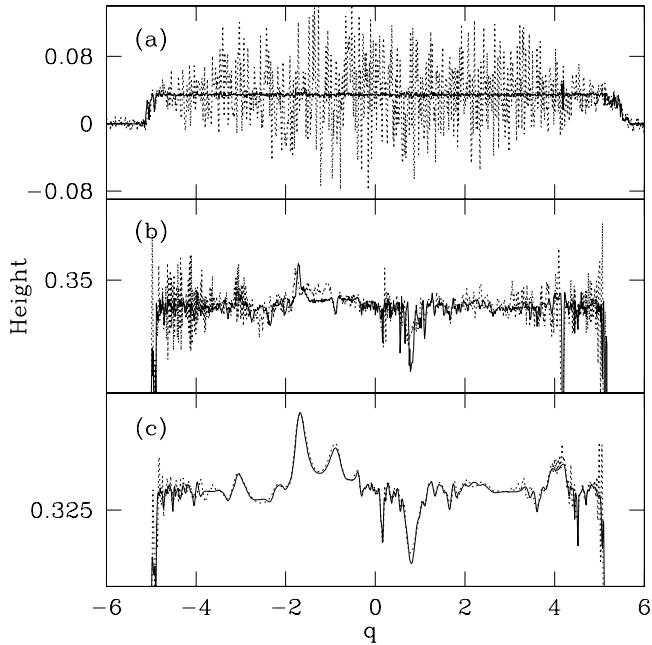}
\caption{\label{fig:SlicePlot} Sectional cuts of Wigner functions
(dashed lines) and classical distributions (solid lines) for a driven
Duffing oscillator, after 149 drive periods, taken at $p=0$ for (a)
$D=10^{-5}$; (b) $D=10^{-3}$; (c) $D=10^{-2}$. Parameter values are as
stated in the text; the height is specified in scaled units.} 
\end{figure}

To illustrate these mechanisms we considered the Duffing oscillator
with unit mass: $H=p^2/2+Bx^4-Ax^2+\Lambda x \cos(\omega t)$.  This
system was studied for a set of parameters ($A=\Lambda=10$, $B=0.5$
and $\omega=6.07$) where the system is strongly chaotic
($\bar{\lambda}=0.57$) ~\cite{Ball} and the dynamical evolution of its
bounded motion is dominated by the homoclinic tangle of a single
hyperbolic fixed point.  As a result, the long-time chaotic evolution
can be completely characterized by the unstable manifold associated
with that fixed point~\cite{Guck}. We chose $\hbar=0.1$ in these
calculations.

The evolution of the Duffing system was calculated for both the
classical and quantum master equations.  Figure~\ref{fig:SlicePlot}
shows sectional cuts at $p=0$ of the quantum and classical phase
distribution functions for three different values of the diffusion
coefficient, $D=10^{-5},~10^{-3},~10^{-2}$, after time $T=149$
evolution periods. As $t^*$ varies slowly with $D$
[Eqns.~(\ref{dist}-\ref{nscale})], in the three cases shown, $t^*$
ranges only from $\sim 20 - 14$ (note that $t^*\ll T$). It is easy to
check that the inequality (\ref{tstar}) is strongly violated for
$D=10^{-5}$, mildly violated for $D=10^{-3}$, and approximately
satisfied for $D=10^{-2}$. For $D=10^{-5}$, the classical and quantum
sections show no similarities, as expected.  The quantum Wigner
function also shows large negative regions, reflecting strong quantum
interference. On increasing $D$ to $10^{-3}$ the magnitude of quantum
coherence decreases dramatically and the classical and quantum slices
have the same average value, as well as specific agreement on some
large scale features.  The two disagree, as expected, on the small
scale structures.  This indicates that, while the quantum and
classical distributions do not exactly match, the Wigner function has
now become sensitive to the larger features of the noise averaged
classical distribution function, indicative of the transition to a
semiclassical regime.  At $D=10^{-2}$, there is near perfect agreement
between classical and quantum distribution functions, save on the
smallest scales.  When $D$ is of order unity, the inequalities
enforcing the QCT at the level of individual
trajectories~\cite{SalKurtTan} are satisfied and the agreement is
essentially exact.  However, as indicated by
Fig.~\ref{fig:SlicePlot}(c), detailed agreement for quantum and
classical distribution functions can begin at much smaller values of
the diffusion constant.
   
\begin{figure}[htbp]
\includegraphics[width=3.5in]{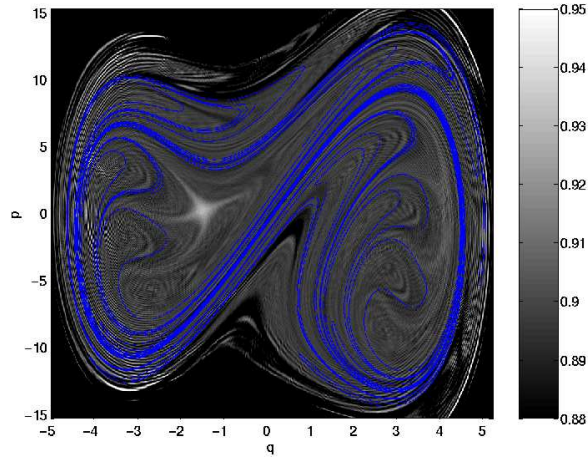}
\caption{\label{fig:WigMan}  Phase space rendering of the Wigner
function at time $t=149~\mbox{periods of driving}$. The early time
part of the unstable manifold associated with the noise-free dynamics
is shown in blue. The value of $D=10^{-3}$ is not sufficient to wipe
out all the quantum interference which, as expected, is most
prominent near turns in the manifold.} 
\end{figure}

For more detailed evidence that, at $D=10^{-3}$, one is entering a
semiclassical regime, in Fig.~\ref{fig:WigMan} we superimpose an image
of the large scale features of the classical unstable manifold on top
of the full quantum Wigner distribution at $D=10^{-3}$ after 149 drive
periods [case (b) of Fig.~1].  The quantum phase space clearly
exhibits local interference fringes around the large lobe-like
structures associated with the short-time evolution of the unstable
manifold.  The appearance of local fringing about classical structures
is direct evidence of a semiclassical evolution, where interference
effects appear locally around the backbone of a classical evolution.
This is in sharp contrast to the global diffraction pattern seen for
$D=0$, where the contributions from individual curves cannot be
distinguished, suppressing the appearance of any classical
structure~\cite{HabibDuff}.

The arguments presented here do not apply directly to systems where
the manifolds are non-compact (analogous to open flows in fluids). In
this case noise does not necessarily lead to a termination of the
structure production in phase space, hence a semiclassical
approximation eventually fails. Thus the topological structure of the
classical manifolds likely provides the physical underpinning of a
proposed classification scheme for dynamical systems based on
quantum-classical correspondence, originally stated in terms of
spectral analysis of the density matrix~\cite{1}. An elaboration of
both the mathematical analysis and numerical results will be presented
in a companion paper~\cite{Prep}.

Numerical simulations were performed on the Cray T3E and IBM SP3 at
NERSC, LBNL. B.D.G. and S.H. acknowledge support from the LDRD program
at LANL. The work of B.S. and B.D.G (partially) was supported by the
National Science Foundation grant \#0099431. This research is supported
by the Department of Energy, under contract W-7405-ENG-36.

\end{document}